\documentclass[%
 reprint,
 amsmath,amssymb,
 aps,
]{revtex4-2}
\newcount\pdfoutput
\pdfoutput=1
\usepackage{amsmath}
\usepackage{graphicx}
\usepackage{subcaption}
\usepackage{dcolumn}
\usepackage{bm}
\usepackage{physics}
\usepackage{tikz-feynman}
\tikzfeynmanset{
every blob={/tikz/fill=white!30,/tikz/inner sep=2pt},
}

\begin{document}

\preprint{APS/123-QED}

\title{On the size of gluon occupancies in saturation}

\author{A.\@ H.\@ Mueller} \email{ahm4@columbia.edu}
\affiliation{%
 Department of Physics \\
 Columbia University \\
 New York, NY 10027
}%

\date{\today}

\begin{abstract}
The size of gluon occupancies, or equivalently the nuclear gluon TMD, at gluon transverse momentum $k_\perp \le Q_s(Y)$ is evaluated. Without Sudakov corrections the occupations can become arbitrarily large while Sudakov effects lead to maximum occupancies of size $(1/\alpha)^{3/2}$. Results are the same for running coupling and fixed coupling dynamics. The coherent (elastic) TMD and inelastic gluon TMD are the same in the gluon saturation region. The saturated gluons in the light cone wavefunction seem to have little or no interaction among themselves.
\end{abstract}

\maketitle


\section{Introduction} \label{sec:1}

In this paper we try to understand how large gluon occupancies can be in the saturation region of a light cone wavefunction. In order to measure such occupancies we consider virtual photon nucleus scattering, using nuclear targets rather than nucleon targets simply because nuclei are more efficient in reaching high levels of saturation. In almost all of the discussion we consider coherent $\gamma^*-A$ scattering where the target nucleus remains in its ground state after the scattering. Essentially all of our conclusions also follow for inelastic collisions, but the issues seem a bit sharper when dealing with coherent scattering. The particular coherent reaction that we shall use is $\gamma^*(q) + A \to q\bar{q}g(k) + A$ where $Q^2 = -q^2$ is very large and, in the rest frame of the nucleus, the $q\bar{q}g$ system is forward moving (along the $\gamma^*$ direction) with the $q$ and $\bar{q}$ having equal and opposite transverse momentum $\underline{p}$ and $-\underline{p}$ with $\underline{p}^2$ on the order of $Q^2$ while the forward softer gluon has a transverse momentum $k_\perp$ less than or equal to the saturation momentum $Q_s(Y)$ of the target and $k_+$ is variable. This reaction has previously been studied \cite{cite1, cite2}, however, here the focus is a little different with our emphasis on the gluon TMD of the nucleus especially the coherent diffractive TMD and how large the gluon occupation can become when the $k_\perp$ of the gluon is less than $Q_s$.

Perhaps the key ingredient in our analysis is in the way we view the initial state in the projectile frame where the nucleus is at rest. In case the nucleus is a simple McLerran-Venugopalan \cite{cite3, cite4, cite5} nucleus the picture of our process can be viewed as in Fig.~\ref{fig:3}. We always suppose the $q\bar{q}$ is compact of size $\Delta x_\perp\sim 1/Q$ while the transverse momentum of the gluon has $k_\perp < Q_s$. Then it is this $q\bar{q}g$ system which scatters on the nucleus. However in higher orders, at the time of scattering, there will be extra gluons in the incoming wavefunction even if we impose that the forward outgoing system be just a $q\bar{q}g$ system. (Recall the case of high energy proton nucleus elastic scattering in the black disc regime. At the time of scattering, and in the rest system of the nucleus, the incoming beam consists of many soft gluons as well as the valence quarks making up the proton. However, they are all part of the proton's wavefunction so in the black disc limit the outgoing state of valence quarks, sea quarks and gluons is exactly the same as the incoming state so only a proton is outgoing.) In our case the incoming system is a $\ket{q\bar{q}g(+)}$ state, that is it is the incoming scattering state (an eigenstate of $P_-$ in terms of light cone operators) corresponding to the original perturbative $q\bar{q}$ and $g$ at $x_+ = -\infty$. In the saturation region the outgoing system will, at $x_+ = +\infty$, be just a perturbative $q\bar{q}g$ system corresponding to
\begin{equation*}
    \bra{q\bar{q}g} S \ket{q\bar{q}g} = \bra{q\bar{q}g(-)}\ket{q\bar{q}g(+)} = 0
\end{equation*}
or
\begin{equation*}
    \bra{q\bar{q}g} T \ket{q\bar{q}g} = 1 \,.
\end{equation*}
Thus in the strong scattering regime, where $k_\perp < Q_s$, unitarity requires that the outgoing forward system be simply the same $q\bar{q}g$ system that came from the $\gamma^*$ initially. This is the key idea in all that follows.

Let me now summarize the different sections of this paper pointing out the main results. In Sec.~\ref{sec:2} we review the McLerran-Venugopalan model (MV) \cite{cite6, cite7, cite8}. The gluon occupancy of the MV nucleus is given by \eqref{eq:5} which is also the gluon TMD for inelastic reactions. (There are no coherent reactions in this simple MV model.) Eq. \ref{eq:5} is surprisingly close to results which will emerge later in this paper even though, except for the measured gluon and the gluon distribution $xG$ in \eqref{eq:4}, all the dynamics in \eqref{eq:5} is effectively classical. Since this model is too simple, and classical, to touch the issue of limits on gluon occupancies we list it for completeness but do not pursue it further.

In Sec.~\ref{sec:3} we introduce our incoming process $\gamma^* \to q\bar{q} \to q\bar{q}g$ in the simple context of scattering on an MV nucleus. In this simple model gluon occupancies, given by \eqref{eq:7}, can become arbitrarily large when the overall rapidity of the process becomes large. What is interesting here is that the elastic scattering of the $q\bar{q}g$ system is limited by unitarity but the growth, $(Y-Y_0)$ in \eqref{eq:7} comes from the various rapidities that the gluon, $k$, can take. This is the essential reason for extremely large occupancies in the more realistic pictures developed in Secs.~\ref{sec:4} and \ref{sec:5}. See also \cite{cite9}.

In Sec.~\ref{sec:4} we consider the same incoming sequence $\gamma^*\to q\bar{q} \to q\bar{q}g$ which scatters on a nucleus but now the nucleus is also evolved by BK \cite{cite10, cite11} evolution. Our answer for the coherent gluon occupancies given in \eqref{eq:8} is almost identical with that of Sec.~\ref{sec:3} with only $Y_0$ changing to $\bar{Y}$ but the essential growth of the TMD's being identical and, at large $Y$, corresponding to arbitrarily large gluon occupancies. In this version of the scattering, unitarity limits are enforced but do not change the linear growth of the occupancies in $Y$ as seen in \eqref{eq:8}. In this section we develop the idea that the incoming scattering state is $\ket{q\bar{q}g(+)}$ and that it is the strong scattering of that state that leads to coherent diffraction and the exact relation \eqref{eq:8} with no undetermined factors.

In Sec.\ref{sec:5} we note that Sudakov effects \cite{cite12, cite13} come in because we require the gluon $k_\perp$ to be less than $Q_s$ in our diffractive TMD. The Sudakov suppression factor occurs because the sequence we follow from $\gamma^*\to q\bar{q}$ and then $q\bar{q}\to q\bar{q}g$ with the gluon $k_\perp < Q_s$ requires that the $q\bar{q}$ not emit higher momentum gluons, $l_\perp$, with $p_\perp > l_\perp > k_\perp$. Such a Sudakov factor is to be expected. After including the Sudakov factor the result \eqref{eq:14} emerges for the gluon occupancy or the gluon TMD. We then argue that $Q^2$ may be chosen as small as $N_0 Q_s^2(Y)$, with $N_0$ some large number, and the $q\bar{q}$ pair still remains a good measurer of the nuclear gluon TMD. This finally leads to a maximum gluon occupancy \eqref{eq:16} at a transverse momentum given by \eqref{eq:17}. Thus gluon occupancies as large as $1/\alpha^{3/2}$ can occur, in either a fixed coupling or running coupling calculation imposing both BK and Sudakov evolution.

In Sec.\ref{sec:6} we note that gluon occupancies or TMD's, for inelastic reactions are identical to those for coherent reactions so long as the gluon transverse momentum is in the saturation regime.

We believe that \eqref{eq:14}, or \eqref{eq:28}, are general results which lead to maximum occupancies of size $(1/\alpha)^{3/2}$. We note that most of the transverse energy in the saturated system is carried by gluons having $k_\perp \sim Q_s$ and occupancies $\sim 1/\alpha$. It is these gluons which dominate the initial stages after a high energy heavy ion collision \cite{cite14, cite15}.

\section{The McLerran-Venugopalan model} \label{sec:2}

We begin our technical discussion of gluon saturation with the McLerran-Venugopalan model in its simplest form using a ``current'' $j=-F_{\mu\nu}^2/4$ to produce a single fast gluon in a current-nucleus scattering. The picture of the process is shown in Fig.~\ref{fig:1}, where the current creates a gluon somewhere in the nucleus and that high momentum gluon can then carry out multiple scatterings as it passes through the MV nucleus. We take the current $j$ to have momentum
\begin{equation} \label{eq:1}
    q = \qty(q_+, -\frac{Q^2}{2q_+},0)
\end{equation}
with $q_+$ large in this projectile frame where the nucleons, $p_i$, are at rest. The multiple scatterings in the amplitude and complex conjugate amplitude lead to a dipole scattering in the nucleus and, with an integration over the longitudinal position of the first scattering, gives the number of produced gluons as \cite{cite7, cite16}
\begin{equation} \label{eq:2}
    \frac{\dd{N}}{\dd[2]{b}\dd[2]{k}} = \int \frac{\dd[2]{x_\perp}}{(2\pi)^2} e^{i \underline{k} \cdot \underline{x}} \tilde{N}(\underline{x},\underline{b})
\end{equation}
where
\begin{equation} \label{eq:3}
    \tilde{N}(\underline{x},\underline{b}) = \frac{N_c^2-1}{\pi^2\alpha N_c \underline{x}^2} \qty(1-e^{-\underline{x}^2 Q_s^2/4}) \,.
\end{equation}

\begin{figure}[htbp]
\centering
\begin{tikzpicture}
\begin{feynman} 
\vertex (q);
\vertex [below right=1.5cm of q] (t1);
\vertex [right=1.5cm of t1] (t2);
\vertex [right=1.5cm of t2] (t3);
\vertex [right=1.5cm of t3] (t4);
\vertex [right=1.5cm of t4] (t5);
\vertex [below=1.5cm of t1, blob, minimum height=0.5cm, minimum width=0.5cm] (b1) {};
\vertex [below=1.5cm of t2, blob, minimum height=0.5cm, minimum width=0.5cm] (b2) {};
\vertex [below=1.5cm of t4, blob, minimum height=0.5cm, minimum width=0.5cm] (b4) {};
\vertex [below left=0.5cm of b1] (l1);
\vertex [right=0.5cm of b1] (r1);\vertex [below left=0.5cm of b2] (l2);
\vertex [right=0.5cm of b2] (r2);\vertex [below left=0.5cm of b4] (l4);
\vertex [right=0.5cm of b4] (r4);

\diagram* {
  (q) -- [scalar, momentum=$q$] (t1) -- [photon, momentum=$q+\delta_1$] (t2) -- [photon, momentum=$q+\delta_2$] (t3) -- [photon] (t4) -- [photon, momentum=$q+\delta_k$] (t5),
  (b1) -- [photon, momentum'=$\delta_1$] (t1),
  (b2) -- [photon, momentum'=$\delta_2-\delta_1$] (t2),(b4) -- [photon, momentum'=$k-\delta_{m-1}$] (t4),
  (l1) -- [double] (b1) -- [double] (r1),
  (l2) -- [double] (b2) -- [double] (r2),
  (l4) -- [double] (b4) -- [double] (r4),
};

\vertex [below left=0.8cm of t4] {$\cdots$};
\vertex [below=0.6cm of b1] {$p_1$};
\vertex [below=0.6cm of b2] {$p_2$};
\vertex [below=0.6cm of b4] {$p_m$};
\end{feynman}
\end{tikzpicture}
\caption{ \label{fig:1}}
\end{figure}

The same result, \eqref{eq:2} and \eqref{eq:3} above, also can be obtained in a target frame where $p_-$ is large and an $A_- = 0$ guage is used \cite{cite16}. The picture now is shown in Fig.~\ref{fig:2} where the gluon fields more forward, in $x_-$, rotate the gluon fields of the gluons having smaller $x_-$ which is equivalent to the $A_+ = 0$ scattering of Fig.~\ref{fig:1} giving \eqref{eq:2} and \eqref{eq:3} above. (In \eqref{eq:3} above
\begin{equation} \label{eq:4}
    Q_s^2 = \frac{8\pi^2\alpha N_c}{N_c^2-1} \sqrt{R^2-b^2} \rho\, xG
\end{equation}
with $\rho$ the nuclear density, $xG$ the gluon distribution of a nucleon in the nucleus, $R$ the radius of the nucleus and $b$ the impact parameter of the incoming current.)

\begin{figure}[htbp]
\centering
\begin{tikzpicture}
\begin{feynman} 
\vertex [blob, minimum height=0.5cm, minimum width=0.5cm] (n1) {};
\vertex [left=0.5cm of n1] (l1);
\vertex [right=0.5cm of n1] (r1);
\vertex [right=3cm of n1, blob, minimum height=0.5cm, minimum width=0.5cm] (n2) {};
\vertex [left=0.5cm of n2] (l2);
\vertex [right=0.5cm of n2] (r2);
\vertex [right=3cm of n2, blob, minimum height=0.5cm, minimum width=0.5cm] (n3) {};
\vertex [left=0.5cm of n3] (l3);
\vertex [right=0.5cm of n3] (r3);
\vertex [below right=2.2cm of n1] (b1);
\vertex [below left=1.12cm of n2] (t1);
\vertex [below left=1.12cm of n3] (t2);
\vertex [right=3cm of b1] (b2);
\vertex [right=1.5cm of b2] (b3);
\vertex [below left=1.5cm of b2] (q);

\diagram* {
  (l1) -- [double] (n1) -- [double] (r1),
  (l2) -- [double] (n2) -- [double] (r2),
  (l3) -- [double] (n3) -- [double] (r3),
  (n1) -- [photon, quarter right] (b1) -- [photon] (b2) -- [photon, momentum'=$q+k$] (b3),
  (b1) -- [photon] (t1) -- [photon] (n2),
  (t1) -- [photon] (t2) -- [photon, quarter right] (n3),
  (q) -- [scalar, momentum'=$q$] (b2),
};

\vertex [above left=0.5cm of t2] {$\cdots$};
\end{feynman}
\end{tikzpicture}
\caption{ \label{fig:2}}
\end{figure}

The scattering given by \eqref{eq:1} and \eqref{eq:2} is strong when $k_\perp/Q_s < 1$ leading to
\begin{equation} \label{eq:5}
    \frac{\dd{N}}{\dd[2]{b} \dd[2]{k}} \simeq \frac{N_c^2 - 1}{4\pi^3 \alpha N_c} \int_{1/Q_s^2}^{1/k_\perp^2} \frac{\dd{x_\perp^2}}{x_\perp^2} = \frac{N_c^2-1}{4\pi^3\alpha N_c} \ln\frac{Q_s^2}{k_\perp^2} \,.
\end{equation}
In the target frame $\frac{\dd{N}}{\dd[2]{b} \dd[2]{k}}$ is the same as the number of gluons per unit impact parameter and per unit transverse momentum in the light cone wavefunction of the nucleus.

Eq.~\eqref{eq:5} shows that the occupation of gluons having $k_\perp$ on the order of $Q_s$ is of size $1/\alpha$ while the occupation of gluons having $k_\perp/Q_s \ll 1$ can be very large. The total number of gluons in the light cone wavefunction of the nucleus is
\begin{equation} \label{eq:6}
    N = \frac{4\pi}{3} R^3 \rho\, xG \,.
\end{equation}
Eq.~\eqref{eq:5} shows that multiple scattering, or gauge rotations in the target picture, pushes most of these gluons to the region $k_\perp \sim Q_s$ where the occupation number is $1/\alpha$ but gluons having $k_\perp \ll 1$ can have extremely high occupancy, and it is this high occupancy which is the focus of this paper.

\section{A single quantum gluon with a MV-target} \label{sec:3}

Now we move a step closer to a realistic circumstance by scattering a highly virtual photon, $\gamma^*$, on a MV-nucleus with the constraint that the $q\bar{q}$ pair coming from the $\gamma^*$ have quark, and antiquark, transverse momentum $p_\perp$ on the order of the $\gamma^*$ virtuality, $Q$. We consider events where the final state consists of the high transverse momentum $q$ and $\bar{q}$ along with a forward gluon, in the projectile frame, having transverse momentum $k_\perp$ which is less than or of the order of the saturation momentum. We focus on coherent events where the nucleus does not break up. The process is illustrated in Fig.~\ref{fig:3} where the elastic scatterings of the $(q\bar{q})g$ ``gluon dipole'' on the target are shown. This process has been discussed in great detail in Refs.~\cite{cite1, cite2} so that here we simply give results. One finds
\begin{equation} \label{eq:7}
    \frac{\dd{N}}{\dd[2]{b} \dd[2]{k}} = \frac{\dd{xG_A}}{\dd[2]{b} \dd[2]{k}} = \frac{N_c^2-1}{8\pi^4} (Y-Y_0)
\end{equation}
where $Y_0$ is the minimum rapidity of the gluon $k+\delta$ so that its lifetime, $\tau_{k+\delta} \simeq 2k_+/k_\perp^2$, is greater than $2R$, the length of the target while $Y$ is the maximum rapidity of the gluon when $2k_+/k_\perp^2 = \tau_k = \tau_\gamma = 2q_+/Q$.

\begin{figure}[htbp]
\centering
\begin{tikzpicture}
\begin{feynman} 
\vertex (p1) {$\gamma$};
\vertex [right=0.7cm of p1] (qa1);
\vertex [above right=0.35cm of qa1] (q1);
\vertex [below right=0.35cm of qa1] (a1);
\vertex [right=0.3cm of qa1] (t1);
\vertex [below right=1.2cm of t1] (b1);
\vertex [right=0.1cm of b1] (b2);
\vertex [right=1.2cm of b2] (b3);
\vertex [right=1.5cm of b3] (b4);
\vertex [right=0.75cm of b3] (bm);
\vertex [above=1.5cm of bm] (tt);
\vertex [below=1.3cm of bm] (bb);
\vertex [right=1.2cm of b4] (b5);
\vertex [right=0.1cm of b5] (b6);
\vertex [above right=1.2cm of b6] (t2);
\vertex [right=0.3cm of t2] (qa2);
\vertex [above left=0.35cm of qa2] (q2);
\vertex [below left=0.35cm of qa2] (a2);
\vertex [right=0.7cm of qa2] (p2) {$\gamma$};
\vertex [above=0.25cm of b2] (nt1);
\vertex [below=0.25cm of b2, blob, minimum height=0.5cm, minimum width=0.5cm] (n1) {};
\vertex [above=0.22cm of n1] (nb1);
\vertex [left=0.1cm of nt1] (ntl1);
\vertex [right=0.1cm of nt1] (ntr1);
\vertex [left=0.1cm of nb1] (nbl1);
\vertex [right=0.1cm of nb1] (nbr1);
\vertex [below left=0.5cm of n1] (nl1);
\vertex [below right=0.5cm of n1] (nr1);
\vertex [above=0.25cm of b3] (nt2);
\vertex [below=0.25cm of b3, blob, minimum height=0.5cm, minimum width=0.5cm] (n2) {};
\vertex [above=0.22cm of n2] (nb2);
\vertex [left=0.1cm of nt2] (ntl2);
\vertex [right=0.1cm of nt2] (ntr2);
\vertex [left=0.1cm of nb2] (nbl2);
\vertex [right=0.1cm of nb2] (nbr2);
\vertex [below left=0.5cm of n2] (nl2);
\vertex [below right=0.5cm of n2] (nr2);
\vertex [above=0.25cm of b4] (nt3);
\vertex [below=0.25cm of b4, blob, minimum height=0.5cm, minimum width=0.5cm] (n3) {};
\vertex [above=0.22cm of n3] (nb3);
\vertex [left=0.1cm of nt3] (ntl3);
\vertex [right=0.1cm of nt3] (ntr3);
\vertex [left=0.1cm of nb3] (nbl3);
\vertex [right=0.1cm of nb3] (nbr3);
\vertex [below left=0.5cm of n3] (nl3);
\vertex [below right=0.5cm of n3] (nr3);
\vertex [above=0.25cm of b5] (nt4);
\vertex [below=0.25cm of b5, blob, minimum height=0.5cm, minimum width=0.5cm] (n4) {};
\vertex [above=0.22cm of n4] (nb4);
\vertex [left=0.1cm of nt4] (ntl4);
\vertex [right=0.1cm of nt4] (ntr4);
\vertex [left=0.1cm of nb4] (nbl4);
\vertex [right=0.1cm of nb4] (nbr4);
\vertex [below left=0.5cm of n4] (nl4);
\vertex [below right=0.5cm of n4] (nr4);
\vertex [right=2.8cm of q1] (qm);
\vertex [above=0.01cm of qm] {$q$};
\vertex [right=2.8cm of a1] (am);
\vertex [below=0.01cm of am] {$\bar{q}$};

\diagram*{
  (p1)  -- [photon] (qa1) -- [quarter left] (q1) -- (q2) -- [quarter left] (qa2) -- [photon] (p2),
  (qa1) -- [quarter right] (a1) -- (a2) -- [quarter right] (qa2),
  (t1) -- [photon, quarter right, momentum'=$k$] (b1) -- [photon] (b3) -- [photon, momentum'=$k+\delta$] (b4) -- [photon] (b6) -- [photon, quarter right] (t2),
  (ntl1) -- [photon, double] (nbl1),
  (ntr1) -- [photon, double] (nbr1),
  (nl1) -- [double] (n1) -- [double] (nr1),
  (ntl2) -- [photon, double] (nbl2),
  (ntr2) -- [photon, double] (nbr2),
  (nl2) -- [double] (n2) -- [double] (nr2),
  (ntl3) -- [photon, double] (nbl3),
  (ntr3) -- [photon, double] (nbr3),
  (nl3) -- [double] (n3) -- [double] (nr3),
  (ntl4) -- [photon, double] (nbl4),
  (ntr4) -- [photon, double] (nbr4),
  (nl4) -- [double] (n4) -- [double] (nr4),
};

\diagram* {
  (tt) -- (bb),
};

\vertex [right=0.63cm of n1] {$\cdots$};
\vertex [right=0.63cm of n3] {$\cdots$};
\end{feynman}
\end{tikzpicture}
\caption{ \label{fig:3}}
\end{figure}

In Fig.~\ref{fig:4} the process is viewed in the target frame with the multiple elastic scatterings of Fig.~\ref{fig:3} grouped into the $\delta$-lines of Fig.~\ref{fig:4}. Fig.~\ref{fig:4} makes clear that at fixed $k_\perp$ this process measures the $k_\perp$-dependent gluon distribution as given in \eqref{eq:7} above.

What is new here is that the scattering process, the $(q\bar{q}g)$ system scattering on the nucleus, is now limited by unitarity. However, while the $(q\bar{q}g)$ system has a limited cross section on the nucleus, for a given $k$, the growth of \eqref{eq:7}, the $Y-Y_0$ term, comes from integrating over the rapidity of the gluon $k$ and this growth is not limited by unitarity. Thus at this level of one quantum gluon in the scattering, gluon occupancies can become arbitrarily large, as given in \eqref{eq:7}, as the total energy, or $Y$, of the process becomes large. This is the heart of the issue we are considering here. That is, are there limits to the gluon occupancies in a nuclear light cone wavefunction when one considers a general QCD scattering? We are now going on to the more general case in order to answer this question.

\begin{figure}[htbp]
\centering
\begin{tikzpicture}
\begin{feynman} 
\vertex [blob, minimum height=0.5cm, minimum width=0.5cm] (n1) {};
\vertex [left=0.5cm of n1] (nl1);
\vertex [right=0.5cm of n1] (nr1);
\vertex [below=1.5cm of n1] (tl);
\vertex [below right=1.5cm of tl] (sul);
\vertex [right=0.5cm of sul] (sur);
\vertex [above right=1.5cm of sur] (tr);
\vertex [right=2.62cm of n1, blob, minimum height=0.5cm, minimum width=0.5cm] (n2) {};
\vertex [left=0.5cm of n2] (nl2);
\vertex [right=0.5cm of n2] (nr2);
\vertex [below=0.5cm of sul] (sbl);
\vertex [right=0.5cm of sbl] (sbr);
\vertex [below left=1cm of sbl] (p1) {$\gamma$};
\vertex [below right=1cm of sbr] (p2) {$\gamma$};
\vertex [right=0.25cm of sul] (sm);
\vertex [above=3cm of sm] (tt);
\vertex [below=2cm of sm] (bb);

\diagram* {
  (nl1) -- [double] (n1) -- [double] (nr1),
  (nl2) -- [double] (n2) -- [double] (nr2),
  (n1) -- [gluon, momentum'=$\delta$] (tl) -- [photon, momentum=$\delta-k$] (tr) -- [gluon, momentum'=$\delta$] (n2),
  (tl) -- [photon, momentum'=$k$] (sul) -- (sur) -- [photon] (tr),
  (sul) -- (sbl) -- (sbr) -- (sur),
  (p1) -- [photon, momentum] (sbl),
  (sbr) -- [photon, momentum] (p2),
};

\diagram* {
  (tt) -- (bb),
};
\end{feynman}
\end{tikzpicture}
\caption{ \label{fig:4}}
\end{figure}

\section{More general target evolution in coherent electron scattering} \label{sec:4}

We here consider the same process as in Sec.~\ref{sec:3} but now we suppose that QCD evolution takes place well beyond a simple MV picture of the nucleus. In particular this means that the saturation momentum depends on the rapidity of the process. In Fig.~\ref{fig:5} we sketch the dependence of the saturation momentum with the rapidity $y$ of the produced forward gluon the line $k+\delta$ in Fig.~\ref{fig:3}. Eq.~\eqref{eq:7} remains valid with the replacement of $Y_0$ by $\bar{Y}$ where $Q_s^2(\bar{Y}) = k_\perp^2$ so that
\begin{equation} \label{eq:8}
    \frac{\dd{N}}{\dd[2]{b} \dd[2]{k}} = \frac{\dd{xG_A}}{\dd[2]{b} \dd[2]{k}} = \frac{N_c^2-1}{8\pi^4} (Y-\bar{Y})
\end{equation}
where now $Y-\bar{Y}$ is the size of the rapidity interval $\bar{Y} < y < Y$ of the gluon $k+\delta$ where the scattering remains strong as is clear from Fig.~\ref{fig:5}. At the moment \eqref{eq:8} is an equation for coherent scattering, as was \eqref{eq:7}, where the outgoing state is a $q\bar{q}$ pair of froward jets and a forward gluon jet having $k_\perp$. We shall shortly argue that \eqref{eq:8} also holds true for inelastic interactions where the nucleus breaks up, but right now is the appropriate time to give a deeper pictures for \eqref{eq:8} in terms of coherent reactions.

\begin{figure}[htbp]
\centering
\begin{tikzpicture}
\begin{feynman}
\vertex (o);
\vertex [right=1cm of o] (x1);
\vertex [right=2cm of x1] (x2);
\vertex [right=0.5cm of x2] (xlarr);
\vertex [right=0.7cm of xlarr] (xrarr);
\vertex [right=2cm of x2] (x3);
\vertex [right=1cm of x3] (x4);
\vertex [above=1cm of o] (y1);
\vertex [above=2cm of y1] (y2);
\vertex [above=1cm of y2] (yarr);
\vertex [above=2cm of y2] (y3);
\vertex [above=1cm of y3] (y4);
\vertex [above right=0.5cm of o] (d0);
\vertex [above=1cm of x1] (d1);
\vertex [above=3cm of x2] (d2);
\vertex [above=5cm of x3] (d3);
\vertex [above=6cm of x4] (d4);
\vertex [right=0.5cm of d2] (lab);

\diagram* {
  (o) -- (x1) -- (x2) -- (xlarr) -- [momentum'] (xrarr) -- (x3) -- (x4),
  (o) -- (y1) -- (y2) -- [momentum] (yarr) -- (y3) -- (y4),
  (d0) -- (d4),
  (y3) -- [scalar] (d3),
  (x3) -- [scalar] (d3),
  (x1) -- [scalar] (d1),
  (y1) -- [scalar] (d1),
  (lab) -- [draw=none, quarter left, momentum] (d2);
};

\vertex [below=0.05cm of x1] {$\ln{k_\perp^2}$};
\vertex [below=0.05cm of x2] {$\ln{Q_s^2(y)}$};
\vertex [below=0.05cm of x3] {$\ln{Q_s^2(Y)}$};
\vertex [left=0.05cm of y1] {$\bar{Y}$};
\vertex [left=0.12cm of y2] {$y$};
\vertex [left=0.05cm of y3] {$Y$};
\vertex [right=0.01cm of lab] {$\ln{Q_s^2(y)}$};
\end{feynman}
\end{tikzpicture}
\caption{ \label{fig:5}}
\end{figure}

Referring to Fig.~\ref{fig:3}, but with the target nucleus now a fully evolved nucleus instead of the MV nucleus shown in Fig.~\ref{fig:3}. One sees that long before the scattering occurs the incoming system consists of a compact $q\bar{q}$ and a relatively soft gluon. In our current general circumstance this $q\bar{q}g$ system further evolves before the scattering. This evolution can be characterized by recognizing that the incoming state $\ket{q\bar{q}g(+)}$ is an incoming scattering state which in $A_+=0$ gauge is an eigenstate of the ``Hamiltonian'' $P_-$ in light cone quantization. At the time of scattering the $\ket{q\bar{q}g(+)}$ state consists of many gluons along with the original $q\bar{q}$ and $g$. However since the scattering is at the unitarity limit, because $k_\perp < Q_s(Y)$, only two types of events are produced. (i) The $q\bar{q}g$ evolved system and the nucleus may break up into a complicated final state inelastic reaction, or (ii) the nucleus remains in its ground state and the $q\bar{q}g$ jets are produced in the forward direction with a large rapidity gap between the $q\bar{q}g$ system and the target nucleus. The coherent and inelastic reactions are of the same size and when interpreted in the target frame in an $A_-=0$ gauge lead to nuclear gluon TMD's given by \eqref{eq:8} with exactly the same values for the coherent and incoherent TMD's.

In the projectile frame where the picture is as in Fig.~\ref{fig:3} this is a reasonably familiar picture. For example in high energy proton-nucleus collisions one has a similar phenomenon where there are again two types of events. (i) Proton-nucleus collisions lead to inelastic reactions where the proton and nucleus both break up or (ii) proton-nucleus collisions can lead to elastic scattering where the nucleus remains in its ground state with the proton and nucleus the only final state particles. For high energies and large nuclei (where edge effects are unimportant) the inelastic and elastic cross sections are each of size $\pi R^2$ with $R$ the nuclear radius.

The picture in the target frame, as seen in Fig.~\ref{fig:4}, is less familiar and does not seem to have a hadronic analog. In this picture the small-$x$ gluon distribution can be very large when $(Y-\bar{Y})$ is large, as seen in \eqref{eq:8}, however these gluons do not seem to interact even though their phase space occupations are very large. In the projectile frame the different $\dd{y}$ regions, adding up to $Y-\bar{Y} = \int_{\bar{Y}}^Y \dd{y}$, are clearly independent as they correspond to different initial $\ket{q\bar{q}g(+)}$ states. However in the target frame the different $\dd{y}$ regions all contribute \textit{independently} to give the gluon TMD of the nucleus. Thus the coherent gluon distribution of the nucleus can become very large because it is \textit{proportional} to the $q\bar{q}$ elastic scattering on the nucleus while unitarity only constrains the $q\bar{q}g$ elastic scattering on the nucleus, so long as the $q\bar{q}$ elastic cross section is small. Thus the various longitudinal momenta ${\dd{k_+}}/k_+ = \dd{y}$ of the incoming gluon all contribute to the gluon TMD of the nucleus but are not strongly constrained by unitarity.

Let's go into this issue of the noninteraction of the small-$x$ gluons in the gluon TMD of the nucleus in even more detail. Consider the graphs of Fig.~\ref{fig:6} in a region where $k_\perp,k_{1\perp}$ are well below $Q_s$. We know from the general theory of scattering that all graphs of the type shown in Fig.~\ref{fig:6A} must cancel with only the $q\bar{q}g(k)$ appearing in the final state when $k_\perp,k_{1\perp} < Q_s$. And, indeed the cancellation of gluon $k_1$ in Fig.~\ref{fig:6A} between its various times of emission (initial state and final state) is well understood \cite{cite17} \footnote{E.\@ Iancu (private correspondence)}. This means that the same must happen for the graphs, like that in Fig.~\ref{fig:6B}, in the target frame. Exactly how this happens is not known, but interactions of the soft gluons (making up saturation) among themselves must cancel so as to leave the contribution of Fig.~\ref{fig:4}, and \eqref{eq:8}, as the complete result agreeing with the projectile frame results.

\begin{figure}[htbp]
\centering
\begin{subfigure}[b]{0.49\textwidth}
\centering
\begin{tikzpicture}
\begin{feynman} 
\vertex (p1) {$\gamma$};
\vertex [right=0.7cm of p1] (qa1);
\vertex [above right=0.35cm of qa1] (q1);
\vertex [below right=0.35cm of qa1] (a1);
\vertex [right=0.3cm of qa1] (t1);
\vertex [below right=1.2cm of t1] (b1);
\vertex [right=0.3cm of b1] (b2);
\vertex [right=0.8cm of b2] (b3);
\vertex [right=1.9cm of b3] (b4);
\vertex [right=0.95cm of b3] (bm);
\vertex [above=0.25cm of b2] (tt1);
\vertex [below right=1.2cm of tt1] (bb1);
\vertex [above=0.25cm of b5] (tt2);
\vertex [below left=1.2cm of tt2] (bb2);
\vertex [above=1.5cm of bm] (tt);
\vertex [below=1.3cm of bm] (bb);
\vertex [right=0.8cm of b4] (b5);
\vertex [right=0.3cm of b5] (b6);
\vertex [above right=1.2cm of b6] (t2);
\vertex [right=0.3cm of t2] (qa2);
\vertex [above left=0.35cm of qa2] (q2);
\vertex [below left=0.35cm of qa2] (a2);
\vertex [right=0.7cm of qa2] (p2) {$\gamma$};
\vertex [above=0.25cm of b3] (nt2);
\vertex [below=1.2cm of b3, blob, minimum height=0.5cm, minimum width=0.5cm] (n1) {A};
\vertex [below left=0.5cm of n1] (nl1);
\vertex [below right=0.5cm of n1] (nr1);
\vertex [above=0.25cm of b4] (nt3);
\vertex [below=1.2cm of b4, blob, minimum height=0.5cm, minimum width=0.5cm] (n2) {A};
\vertex [below left=0.5cm of n2] (nl2);
\vertex [below right=0.5cm of n2] (nr2);
\vertex [right=2.8cm of q1] (qm);
\vertex [right=2.8cm of a1] (am);

\diagram* {
  (p1)  -- [photon] (qa1) -- [quarter left] (q1) -- (q2) -- [quarter left] (qa2) -- [photon] (p2),
  (qa1) -- [quarter right] (a1) -- (a2) -- [quarter right] (qa2),
  (t1) -- [photon, quarter right, momentum'=$k$] (b1) -- [photon] (b2) -- [photon] (b3) -- [photon] (b4) -- [photon] (b5) -- [photon] (b6) -- [photon, quarter right] (t2),
  (tt1) -- [photon, quarter right, momentum'=$k_1$] (bb1) -- [photon] (bb2) -- [photon, quarter right] (tt2),
  (nl1) -- [double] (n1) -- [double] (nr1),
  (nl2) -- [double] (n2) -- [double] (nr2),
};

\diagram* {
  (tt) -- (bb),
};
\end{feynman}
\end{tikzpicture}
\caption{\label{fig:6A}}
\end{subfigure}
\hfill
\vspace{0.5cm}
\begin{subfigure}[b]{0.49\textwidth}
\centering
\begin{tikzpicture}
\begin{feynman}
\vertex [blob, minimum height=0.5cm, minimum width=0.5cm] (n1) {$A$};
\vertex [left=0.5cm of n1] (nl1);
\vertex [right=0.5cm of n1] (nr1);
\vertex [below=1cm of n1] (tl);
\vertex [below right=0.5cm of tl] (l);
\vertex [below right=1cm of l] (sul);
\vertex [right=0.5cm of sul] (sur);
\vertex [above right=1cm of sur] (r);
\vertex [above right=0.5cm of r] (tr);
\vertex [right=2.62cm of n1, blob, minimum height=0.5cm, minimum width=0.5cm] (n2) {$A$};
\vertex [left=0.5cm of n2] (nl2);
\vertex [right=0.5cm of n2] (nr2);
\vertex [below=0.5cm of sul] (sbl);
\vertex [right=0.5cm of sbl] (sbr);
\vertex [below=0.5cm of sbl] (p1) {$\gamma$};
\vertex [below=0.5cm of sbr] (p2) {$\gamma$};
\vertex [below left=1.5cm of sbl] (bl);
\vertex [below right=1.5cm of sbr] (br);
\vertex [below=4.621cm of n1, blob, minimum height=0.5cm, minimum width=0.5cm] (n3) {$A$};
\vertex [left=0.5cm of n3] (nl3);
\vertex [right=0.5cm of n3] (nr3);
\vertex [below=4.621cm of n2, blob, minimum height=0.5cm, minimum width=0.5cm] (n4) {$A$};
\vertex [left=0.5cm of n4] (nl4);
\vertex [right=0.5cm of n4] (nr4);
\vertex [right=0.25cm of sul] (sm);
\vertex [above=2.5cm of sm] (tt);
\vertex [below=3cm of sm] (bb);

\diagram* {
  (nl1) -- [double] (n1) -- [double] (nr1),
  (nl2) -- [double] (n2) -- [double] (nr2),
  (nl3) -- [double] (n3) -- [double] (nr3),
  (nl4) -- [double] (n4) -- [double] (nr4),
  (n1) -- [gluon] (tl) -- [photon, momentum=$k_2$] (tr) -- [gluon] (n2),
  (tl) -- [photon] (l) -- [photon, momentum=$k$] (sul) -- (sur) -- [photon, momentum=$k$] (r) -- [photon] (tr),
  (sul) -- (sbl) -- (sbr) -- (sur),
  (l) -- [photon] (bl) -- [photon] (br) -- [photon] (r),
  (bl) -- [gluon] (n3),
  (br) -- [gluon] (n4),
  (p1) -- [photon, momentum] (sbl),
  (sbr) -- [photon, momentum] (p2),
};

\diagram* {
  (tt) -- (bb),
};
\end{feynman}
\end{tikzpicture}
\caption{\label{fig:6B}}
\end{subfigure}
\caption{ \label{fig:6}}
\end{figure}

\section{Including Sudakov effects} \label{sec:5}

\subsection{Sudakov effects; fixed coupling} \label{sec:5.1}

As we have just discussed in the previous section of this paper, Equation \eqref{eq:8} follows from the integration of the rapidity $y$ of the produced gluon over the region, shown in Fig.~\ref{fig:5}, $\bar{Y} < y < Y$ where the scattering is strong and the scattering of the state $\ket{q\bar{q}g(+)}$ is at the unitarity limit. We have said earlier that emissions, say $k'$, later in time (in the projectile frame) than the $k$ emission and also having $k'_\perp < Q_s$ cancel. However, an emission of a gluon of momentum $l$ with $l_\perp > Q_s$ earlier than the $k$-emission does not cancel. We do not want such emissions since they would correspond to a TMD at $l_\perp$ rather than $k_\perp$ and to avoid such emissions one must pay the price of a Sudakov suppression, illustrated in Fig.~\ref{fig:7} in the target frame \cite{cite12, cite13, cite19, cite20}.

The lowest order Sudakov contribution coming from Fig.~\ref{fig:7} and a similar graph where the self-energy is in the complex conjugate amplitude is given as
\begin{equation} \label{eq:9}
\begin{split}
    \mathit{Sud}^{(1)} &= -\frac{2\pi g^2N_c}{(2\pi)^3} \int_{k_\perp^2}^{Q^2} \frac{\dd{l_\perp^2}}{l_\perp^2} \int_{l_\perp^2/q_+}^{k_-} \frac{\dd{l_-}}{l_-} \\
    &= -\frac{\alpha N_c}{2\pi} \ln^2\frac{Q^2}{k_\perp^2} \,.
\end{split}
\end{equation}
As far as the limits of the $l$-integration in \eqref{eq:9} are concerned: (i) The upper limit $l_\perp^2 = Q^2$ is set by renormalization. (ii) If $l_\perp^2 < k_\perp^2$ there is a corresponding real $l$-emission which cancels the graph of Fig.~\ref{fig:7}. (iii) $l_- < k_-$ is manifest. (iv) $l_- > l_\perp^2/q_+$ is the requirement that the time of emission of the $l$-line, $2l_-/l_\perp^2$ in the target frame, be larger than the lifetime of the quark loop.

\begin{figure}[htbp]
\centering
\begin{tikzpicture}
\begin{feynman}
\vertex [blob, minimum height=0.5cm, minimum width=0.5cm] (n1) {};
\vertex [left=0.5cm of n1] (nl1);
\vertex [right=0.5cm of n1] (nr1);
\vertex [below=1.5cm of n1] (tl);
\vertex [below right=0.5cm of tl] (l1);
\vertex [below right=1.5cm of l1] (l2);
\vertex [below right=0.5cm of l2] (sul);
\vertex [right=0.5cm of sul] (sur);
\vertex [above right=2.5cm of sur] (tr);
\vertex [right=4.036cm of n1, blob, minimum height=0.5cm, minimum width=0.5cm] (n2) {};
\vertex [left=0.5cm of n2] (nl2);
\vertex [right=0.5cm of n2] (nr2);
\vertex [below=0.5cm of sul] (sbl);
\vertex [right=0.5cm of sbl] (sbr);
\vertex [below left=1cm of sbl] (p1) {$\gamma$};
\vertex [below right=1cm of sbr] (p2) {$\gamma$};
\vertex [right=0.25cm of sul] (sm);
\vertex [above=3cm of sm] (tt);
\vertex [below=2cm of sm] (bb);

\diagram* {
  (nl1) -- [double] (n1) -- [double] (nr1),
  (nl2) -- [double] (n2) -- [double] (nr2),
  (n1) -- [gluon] (tl) -- [photon] (tr) -- [gluon] (n2),
  (tl) -- [photon] (l1) -- [photon, momentum=$k-l$] (l2) -- [photon] (sul) -- (sur) -- [photon] (tr),
  (sul) -- (sbl) -- (sbr) -- (sur),
  (l1) -- [photon, half right, momentum'=$l$] (l2),
  (p1) -- [photon, momentum] (sbl),
  (sbr) -- [photon, momentum] (p2),
};

\diagram* {
  (tt) -- (bb),
};
\end{feynman}
\end{tikzpicture}
\caption{ \label{fig:7}}
\end{figure}

Thus the overall Sudakov factor is
\begin{equation} \label{eq:10}
    \mathit{Sud} = e^{-\frac{\alpha N_c}{2\pi} \ln^2\frac{Q^2}{k_\perp^2}} \,.
\end{equation}
Now let's put \eqref{eq:10} together with \eqref{eq:8}. In order to eliminate the $Y-\bar{Y}$ factor it is useful to recall that \cite{cite8}
\begin{equation} \label{eq:11}
    Q_s^2(Y) = Q_s^2(\bar{Y}) e^{\frac{2\alpha N_c}{\pi} \frac{\chi(\lambda_0)}{1-\lambda_0} (Y-\bar{Y})}
\end{equation}
or
\begin{equation} \label{eq:12}
    Y-\bar{Y} = \frac{1-\lambda_0}{2\chi(\lambda_0)} \frac{1}{\alpha N_c/\pi} \ln\frac{Q_s^2(Y)}{Q_s^2(\bar{Y})}
\end{equation}
which is the same as
\begin{equation} \label{eq:13}
    Y-\bar{Y} = \frac{1-\lambda_0}{2\chi(\lambda_0)} \frac{1}{\alpha N_c/\pi} \ln\frac{Q_s^2(Y)}{k_\perp^2} \,.
\end{equation}
Thus including (fixed coupling) Sudakov factors with BFKL-BK evolution one has
\begin{multline} \label{eq:14}
    \frac{\dd{N}}{\dd[2]{b} \dd[2]{k}} = \frac{\dd{xG_A}}{\dd[2]{b} \dd[2]{k}} \\
    = \frac{N_c^2-1}{8\pi^4} \frac{1-\lambda_0}{2\chi(\lambda_0)} \frac{1}{\alpha N_c/\pi} \ln\frac{Q_s^2(Y)}{k_\perp^2} e^{-\frac{\alpha N_c}{2\pi} \ln^2\frac{Q^2}{k_\perp^2}} \,.
\end{multline}

From \eqref{eq:8} and \eqref{eq:14} one sees that gluon occupations can become much larger than $1/\alpha$ when $\ln(Q_s^2(Y)/k_\perp^2)$ is also large. When Sudakov effects are included, and using $Q^2 > Q_s(Y)$ for our whole discussion to make sense, one still gets occupations much larger than $1/\alpha$ so long as $Q^2$ is not too large. The exact value of $Q^2$ is partially our choice. In order that the quark loop make a measurement of individual gluons in the saturation region one must take $Q_s^2(Y)/Q^2 \ll 1$, but otherwise the exact value of $Q^2$ is our choice. It is convenient to take $Q^2 = N_0 Q_s^2(Y)$ where $N_0$ is a large number but not dependent on $\alpha$. Then the quark loop, say in Fig.~\ref{fig:4}, is a good measurer of single gluons up to inverse powers of $N_0$. Thus in the following we replace the $Q^2$ in \eqref{eq:14} by $Q_s^2(Y)$ which should be fine for the parton interpretation we are considering. Values of $Q^2$ larger than $N_0 Q_s^2(Y)$ simply enhance Sudakov effects by requiring that our measured gluons have no additional gluon clouds down to sizes much smaller than $1/Q_s(Y)$ and this is not a physically reasonable requirement.

Now let's take \eqref{eq:14}, with $Q^2 = Q_s^2(Y)$ and see how large gluon occupancies can become before a Sudakov suppression sets in. When $k_\perp^2/Q_s^2$ is small but not too small then \eqref{eq:14} gives occupancies
\begin{equation} \label{eq:15}
    \frac{\dd{N}}{\dd[2]{b} \dd[2]{k_\perp}} = \frac{N_c^2-1}{8\pi^4} \frac{1-\lambda_0}{2\chi(\lambda_0)} \frac{1}{\alpha N_c/\pi} \ln\frac{Q_s^2(Y)}{k_\perp^2} \,.
\end{equation}
\eqref{eq:15} is the same as (64) of \cite{cite9} except for a factor of $1/2$ reflecting the fact that \eqref{eq:15} is only for coherent reactions. $\frac{\dd{N}}{\dd[2]{b} \dd[2]{k_\perp}}$ reaches a maximum when $\ln(Q_s^2/k_\perp^2) = \sqrt{\pi/\alpha N_c}$ at which value one gets from \eqref{eq:14}
\begin{equation} \label{eq:16}
    \frac{\dd{N}}{\dd[2]{b} \dd[2]{k_\perp}} \simeq \frac{N_c^2-1}{8\pi^4} \frac{1-\lambda_0}{2\chi(\lambda_0)} \qty(\frac{\pi}{\alpha N_c})^{3/2} \frac{1}{\sqrt{e}} \,.
\end{equation}
Thus the maximum occupancies occur at
\begin{equation} \label{eq:17}
    k_\perp^2 \simeq Q_s^2(Y) e^{-\sqrt{\frac{\pi}{2\alpha N_c}}}
\end{equation}
with occupancy on the order of $\alpha^{-3/2}$. Of course there is very little phase space near the $k_\perp^2$ in \eqref{eq:17} so the great majority of saturated gluons are around $k_\perp^2 = Q_s^2(Y)$.

\subsection{Sudakov effects; running coupling} \label{sec:5.2}

We are briefly going to redo the discussion of Sec.~\ref{sec:5.1} but now using a running coupling for both BFKL and for Sudakov effects. We continue to focus on coherent reactions with incoherent reactions to be discussed later in this note.

We start with the lowest order Sudakov term in \eqref{eq:9} but now we use running coupling
\begin{equation} \label{eq:18}
    \alpha(l_\perp^2) = \frac{1}{b \ln(l_\perp^2/\Lambda^2)}
\end{equation}
where $b = (11N_c-2N_g)/12\pi$ so that \eqref{eq:9} becomes
\begin{equation} \label{eq:19}
    \mathit{Sud}^{(1)} = -\frac{N_c}{\pi b} \int_{k_\perp^2}^{Q^2} \frac{\dd{l_\perp^2}}{l_\perp^2} \frac{1}{\ln(l_\perp^2/\Lambda^2)} \int_{l_\perp^2/q_+}^{k_-} \frac{\dd{l_-}}{l_-}
\end{equation}
Eq.~\eqref{eq:19} gives
\begin{equation} \label{eq:20}
    \mathit{Sud}^{(1)} = -\frac{N_c}{\pi b} \qty[\ln\frac{Q^2}{\Lambda^2} \ln(\frac{\ln(Q^2/\Lambda^2)}{\ln(k_\perp^2/\Lambda^2)}) - \ln\frac{Q^2}{k_\perp^2}]
\end{equation}
which can be written as
\begin{equation} \label{eq:21}
    \mathit{Sud}^{(1)} = -\frac{N_c}{\pi b} \qty[\ln\frac{Q^2}{\Lambda^2} \ln(1+\frac{\ln(Q^2/k_\perp^2)}{\ln(k_\perp^2/\Lambda^2)}) - \ln\frac{Q^2}{k_\perp^2}]
\end{equation}
which we will use just below.

Also, in the running coupling case we should use \cite{cite21}
\begin{equation} \label{eq:22}
    \ln\frac{Q_s^2(Y)}{\Lambda^2} \simeq \sqrt{\frac{4N_c}{\pi b} \frac{\chi(\lambda_0)}{1-\lambda_0} Y}
\end{equation}
rather than \eqref{eq:11}. With running coupling the picture in Fig.~\ref{fig:5} changes to that of Fig.~\ref{fig:8}. When $y$ goes between $\bar{Y}$ and $Y$ the saturation momentum goes between $k_\perp^2$ and $Q_s^2(Y)$. We take, as in the fixed coupling case, $Q^2$ somewhat bigger than $Q_s^2(Y)$. In order to avoid a strong Sudakov suppression we take $\frac{\ln(Q^2/k_\perp^2)}{\ln(k_\perp^2/\Lambda^2)} \ll 1$ in \eqref{eq:21} so that formula then reads
\begin{equation} \label{eq:23}
    \mathit{Sud}^{(1)} \simeq -\frac{N_c}{2\pi b} \frac{\ln[2](Q^2/k_\perp^2)}{\ln(k_\perp^2/\Lambda^2)} \simeq -\frac{N_c}{2\pi b} \frac{\ln[2](Q_s^2(Y)/k_\perp^2)}{\ln(k_\perp^2/\Lambda^2)}
\end{equation}
Exponentiating \eqref{eq:23} we get the Sudakov factor
\begin{equation} \label{eq:24}
    \mathit{Sud} = e^{-\frac{N_c}{2\pi b} \frac{\ln[2](Q_s^2(Y)/k_\perp^2)}{\ln(k_\perp^2/\Lambda^2)}}
\end{equation}

Now we go back to \eqref{eq:8} and use \eqref{eq:24} to include the Sudaov factor giving
\begin{equation} \label{eq:25}
    \frac{\dd{N}}{\dd[2]{b} \dd[2]{k_\perp}} = \frac{N_c^2-1}{8\pi^4} (Y-\bar{Y}) e^{-\frac{N_c}{2\pi b} \frac{\ln[2](Q_s^2(Y)/k_\perp^2)}{\ln(k_\perp^2/\Lambda^2)}} \,.
\end{equation}
Finally use \eqref{eq:22} to evaluate $Y-\bar{Y}$ in \eqref{eq:25}. One gets
\begin{equation*}
    Y-\bar{Y} = \qty[\frac{4N_c}{\pi b} \frac{\chi(\lambda_0)}{1-\lambda_0}]^{-1} \qty(\ln^2\frac{Q_s^2(Y)}{\Lambda^2} - \ln^2\frac{k_\perp^2}{\Lambda^2})
\end{equation*}
or
\begin{equation} \label{eq:26}
    Y-\bar{Y} = \qty[\frac{4N_c}{\pi b} \frac{\chi(\lambda_0)}{1-\lambda_0}]^{-1} 2\ln\frac{Q_s^2(Y)}{k_\perp^2} \ln\frac{Q_s^2(Y)}{\Lambda^2}
\end{equation}
Using \eqref{eq:26} in \eqref{eq:25} we get
\begin{widetext}
\begin{equation} \label{eq:27}
    \frac{\dd{N}}{\dd[2]{b} \dd[2]{k_\perp}} = \qty[\frac{2N_c}{\pi b} \frac{\chi(\lambda_0)}{1-\lambda_0}]^{-1} \frac{N_c^2-1}{8\pi^4} \ln\frac{Q_s^2(Y)}{\Lambda^2} \ln\frac{Q_s^2(Y)}{k_\perp^2} e^{-\frac{N_c}{2\pi b} \frac{\ln[2](Q_s^2(Y)/k_\perp^2)}{\ln(Q_s^2(Y)/\Lambda^2)}}
\end{equation}
\end{widetext}
Using the explicit formula, \eqref{eq:18}, for the running coupling one can write \eqref{eq:27} as
\begin{equation} \label{eq:28}
    \frac{\dd{N}}{\dd[2]{b} \dd[2]{k_\perp}} = \frac{N_c^2-1}{8\pi^4} \frac{1-\lambda_0}{2\chi(\lambda_0)} \frac{1}{\alpha N_c/\pi} \ln\frac{Q_s^2(Y)}{k_\perp^2} e^{-\frac{\alpha N_c}{2\pi} \ln^2\frac{Q_s^2(Y)}{k_\perp^2}}
\end{equation}
which is identical to \eqref{eq:14} with the identification of $Q_s^2(Y)$ and $Q^2$. The $\alpha$ in \eqref{eq:28} is $\alpha(Q_s^2(Y))$. Thus the fixed coupling and running coupling calculations give the same result.

\section{Other issues} \label{sec:6}

In this section we briefly discuss inelastic reactions as well as the role of $Q^2$, in particular as it affects Sudakov corrections.

\subsection{Inelastic TMD's and gluon distributions} \label{sec:6.1}

So far we have only considered coherent $\gamma^*$-nucleus scattering and the associated coherent gluon distributions. The generalization to inelastic reactions is straightforward. As an example consider the coherent reaction of Fig.~\ref{fig:3}. As we have discussed earlier one may view the process as the scattering of the state $\ket{q\bar{q}g(+)}$ on a nucleus. (In Fig.~\ref{fig:3} the nucleus is a simple MV nucleus, but as we have seen earlier there is not a lot of difference between a fully evolved QCD nucleus and a MV nucleus.) So far we have considered the elastic scattering of $\ket{q\bar{q}g(+)}$ on the nucleus, but in the strong scattering regime we know that in general the coherent scattering and the inelastic reactions are identical which, in the current circumstance means that the coherent occupation given by \eqref{eq:8} is the same as the corresponding incoherent occupation. Similarly Sudakov effects for elastic and inelastic TMD's are the same.

\subsection{The role of $Q^2$} \label{sec:6.2}

In the scattering process, in the projectile frame, the $q\bar{q}$ system is required to be of transverse size $1/Q$. That is we choose events not in the aligned jet region. The overall cross section is of size $1/Q^2$, however, our interest is not in the cross section but in the gluon TMD of the target. Any $Q^2$ significantly greater than $Q_s^2(Y)$ allows the $q\bar{q}$ pair, in the target frame to be a good measurer of a gluon of transverse momentum $k_\perp < Q_s(Y)$. Eq.~\eqref{eq:14} shows that the gluon transverse momentum dependence of the target TMD depends strongly on the value of $Q^2$ when $\alpha \ln[2](Q^2/k_\perp^2)$ is greater than 1. This is very natural. When $Q^2$ is very large it is a strong requirement to demand that there be no gluon emissions having transverse momentum between $k_\perp$ and $Q$. On the other hand it is important that $Q$ be greater than $Q_s(Y)$ otherwise our whole picture of the projectile frame scattering will not be that of a $\ket{q\bar{q}g(+)}$ state scattering on a target but, if $Q<Q_s(Y)$, that of a $\ket{q\bar{q}(+)}$ state scattering on a target. In our detailed discussion of Sudakov efects we have taken $Q^2 \gg Q_s^2(Y)$ in order that the $q\bar{q}$ coming from the $\gamma^*$ be a good probe of the target gluon distribution, but $Q^2/Q_s^2(Y)$ not too large in order to minimize Sudakov effects.

\subsection{The nuclear wavefunction as a coherent state} \label{sec:6.3}

In an earlier paper \cite{cite22} it was noted that in the small $x$ saturation region the nuclear wavefunction was a coherent state of quark$(k_\perp)$-antiquark$(-k_\perp)$ pairs in a color singlet state when $k_\perp < Q_s$. We now note that the same is true of gluons with the difference that the two gluons, in the target frame, can have widely differing $k_-$ momentum along with equal and opposite transverse momenta in the saturation regime. Here also the two gluons are in a color singlet state.

\begin{figure}[htbp]
\centering
\begin{tikzpicture}
\begin{feynman}
\vertex (o);
\vertex [right=1cm of o] (x1);
\vertex [right=2cm of x1] (x2);
\vertex [right=0.5cm of x2] (xlarr);
\vertex [right=0.7cm of xlarr] (xrarr);
\vertex [right=2cm of x2] (x3);
\vertex [right=1cm of x3] (x4);
\vertex [above=1cm of o] (y1);
\vertex [above=2cm of y1] (y2);
\vertex [above=1cm of y2] (yarr);
\vertex [above=2cm of y2] (y3);
\vertex [above=1cm of y3] (y4);
\vertex [right=0.5cm of o] (foo);
\vertex [above=0.9cm of foo] (d0);
\vertex [above=1cm of x1] (d1);
\vertex [below=0.3cm of y2] (baz);
\vertex [right=4cm of baz] (d2);
\vertex [above=5cm of x3] (d3);
\vertex [above=0.5cm of d3] (bar);
\vertex [right=0.1cm of bar] (d4);
\vertex [left=1cm of d2] (lab);

\diagram* {
  (o) -- (x1) -- (x2) -- (xlarr) -- (xrarr) -- (x3) -- (x4),
  (o) -- (y1) -- (y2) -- [momentum] (yarr) -- (y3) -- (y4),
  (d0) -- [out=15, in=-160] (d1) -- [out=20, in=-110] (d3) -- [out=70, in=-105] (d4),
  (y3) -- [scalar] (d3),
  (x3) -- [scalar] (d3),
  (x1) -- [scalar] (d1),
  (y1) -- [scalar] (d1),
  (lab) -- [draw=none, momentum] (d2);
};

\vertex [below=0.05cm of x1] {$\ln{\dfrac{k_\perp^2}{\Lambda^2}}$};
\vertex [below=0.05cm of x3] {$\ln{\dfrac{Q_s^2(Y)}{\Lambda^2}}$};
\vertex [left=0.05cm of y1] {$\bar{Y}$};
\vertex [left=0.12cm of y2] {$y$};
\vertex [left=0.05cm of y3] {$Y$};
\vertex [above=0.35cm of lab] (fiz);
\vertex [left=0.02cm of fiz] {$\ln{\dfrac{Q_s^2(y)}{\Lambda^2}}$};
\end{feynman}
\end{tikzpicture}
\caption{ \label{fig:8}}
\end{figure}

\begin{acknowledgments}
I have benefited from discussions with Edmond Iancu, Raju Venugopalan and Feng Yuan during this work. This work was supported in part through a DOE grant DE-SC0011941.
\end{acknowledgments}

\nocite{*}

\bibliography{refs}

\end{document}